# Generation of Intense Phase-Stable Femtosecond Hard X-ray Pulse Pairs

Yu Zhang[1,*], Thomas Kroll[2], Clemens Weninger[3], Yurina Michine[4], Franklin D. Fuller[3], Diling Zhu[3], Roberto Alonso-Mori[3], Dimosthenis Sokaras[2], Alberto Lutman[5], Aliaksei Halavanau[6], Claudio Pellegrini[6], Andrei Benediktovitch[7], Makina Yabashi[8,9], Ichiro Inoue[8], Yuichi Inubushi[8,9], Taito Osaka[8], Jumpei Yamada[8], Ganguli Babu[10], Devashish Salpekar[10], Farheen N. Sayed[10], Pulickel M. Ajayan[10], Jan Kern[11], Junko Yano[11], Vittal K. Yachandra[11], Hitoki Yoneda[4], Nina Rohringer[7,12,*], Uwe Bergmann[1,13,*]

[1]Stanford PULSE Institute, SLAC National Accelerator Laboratory; Menlo Park, USA

[2]SSRL, SLAC National Accelerator Laboratory; Menlo Park, USA

[3] LCLS, SLAC National Accelerator Laboratory; Menlo Park, USA

[4] The University of Electro-Communications; Chofu, Tokyo, Japan

[5]Linac & FEL division, SLAC National Accelerator Laboratory; Menlo Park, USA

[6]Accelerator Research Division, SLAC National Accelerator Laboratory; Menlo Park, USA

[7] DESY, Center for Free-Electron Laser Science; Hamburg, Germany

[8] RIKEN SPring-8 Center; 1-1-1 Kouto, Sayo, Hyogo 679-5148, Japan

[9] Japan Synchrotron Radiation Research Institute; 1-1-1 Kouto, Sayo, Hyogo 679-5198, Japan

[10] Department of Materials Science and NanoEngineering; Rice University, Houston, USA

[11] Molecular Biophysics and Integrated Bioimaging Division, Lawrence Berkeley National Laboratory Berkeley, USA

[12] Universität Hamburg, Department of Physics; Hamburg, Germany

[13] Department of Physics, University of Wisconsin–Madison; Madison, USA

* Corresponding author. Email: yu.spacezhang@gmail.com (Y.Z.); nina.rohringer@desy.de (N.R.); ubergmann@wisc.edu (U.B.)

**Abstract:**

**Coherent nonlinear spectroscopies and imaging in the X-ray domain provide direct insight into the coupled motions of electrons and nuclei with resolution on the electronic length and time scale. The experimental realization of such techniques will strongly benefit from access to intense, coherent pairs of femtosecond X-ray pulses. We have observed phase-stable X-ray pulse pairs containing more than $3*10^7$ photons at 5.9 keV (2.1 Å) with ~1 fs duration and 2-5 fs separation. The highly directional pulse pairs are manifested by interference fringes in the superfluorescent and seeded stimulated manganese Kα emission induced by an X-ray free-electron laser. The fringes constitute the time-frequency X-ray analogue of Young's double-slit interference allowing for frequency-domain X-ray measurements with attosecond time resolution.**

**Significance**

The generation of phase-stable femtosecond X-ray pulse pairs will advance nonlinear spectroscopies and imaging providing direct insight into the coupled motions of electrons and nuclei with resolution on the electronic length and time scale. The paper presents the generation of such pulse pairs in the x-ray domain. The approach uses X-ray free-electron laser pulses to induce highly directional, intense, phase-stable pairs of superfluorescence and seeded stimulated emission at the 5.9 keV manganese $K\alpha_1$ line. The finding is evidenced by strong interference fringes in the superfluorescence and stimulated emission signals.

Nonlinear coherent imaging and spectroscopy techniques have revolutionized our understanding of the structures and dynamics of molecules and materials (*1-3*). Mukamel and others (*4, 5*) have proposed the extension of nonlinear optical techniques to the X-ray spectral domain to exploit the advantages of atomic spatial resolution and element sensitivity by core-level excitations. The development of powerful XFELs have enabled new classes of experiments with unprecedented spatial resolution and femtosecond temporal resolution, by various techniques (*6-9*), but the experimental realization of many of the proposed nonlinear X-ray techniques, such as coherent X-ray-pump/X-ray-probe experiments, is very challenging for lack of intense, coherent, femtosecond X-ray pulses with fixed relative phases.

The standard operation of XFELs results in self-amplified spontaneous emission (SASE) (*10*) pulses consisting of many random spectral and temporal spikes with limited longitudinal coherence (with coherence times for hard x-rays pulses in the sub fs range). Self-seeding schemes (*11*) provide monochromatic XFEL pulses with increased temporal coherence, and several groups are pursuing the creation and detection of XFEL pulse pairs (*12-14*). However, no phase stabilized femtosecond hard X-ray pulse pairs have been created to date. A different approach for creating intense coherent X-ray pulses is by collective spontaneous emission (*15-19*) and seeded stimulated emission (*17, 20*), which have been observed and explored in various systems at X-ray energies ranging from 850 eV to 8 keV. In both cases, a SASE XFEL pump pulse creates core-electron excitation of a long, quasi one-dimensional medium in a traveling wave geometry. X-ray fluorescence photons spontaneously emitted in the entrance region along the XFEL propagation direction initiate the collective spontaneous emission along this direction (see Fig. 1). In the initial stages of the process this leads to amplified spontaneous emission (ASE), and when the collective emission becomes strong enough to overcome the decoherence rate of spontaneous emission and Auger decay, superfluorescence emerges (*21-22*). The principles and applications of these inner-shell X-ray lasing phenomena are explored for spectroscopy (*18, 20*) and as a new X-ray source (*15, 23, 24*). If an XFEL SASE pump pulse contains two strong temporal spikes, these can generate two superfluorescence or seeded stimulated emission pulses that are separated by a few fs. In this report we present the experimental evidence and theoretical description of the creation of such phase-stable X-ray pulse pairs. Evidence of these pulse pairs is provides by the observation of interference fringes in superfluorescence and seeded stimulated emission bursts of the manganese Kα fluorescence at 5.9 keV (2.1 Å).

**Figure 1 here**

**Results**

The experiments were performed at the nanofocus instrument EH5 on beamline 3 at the SACLA XFEL providing highly focused SASE pump and seed pulses (see SM). Spectral analysis of the emission signal was performed using a flat Si (220) analyzer crystal dispersing the emission signal onto a two-dimensional CCD with the spectral axis in the vertical direction and the spatial axis in the horizontal direction in a geometry similar to previous experiments (*18, 20*) (see SM).

In Fig. 2, we illustrate the experimental conditions for observing interference fringes. An XFEL SASE pump pulse with two strong temporal spikes (a), (b), impinges on the sample (step 1), each creating a short superfluorescence burst. The two pulses leave the sample with a slight delay with respect to their respective SASE spikes given by the lifetime of the excited state (step 2). The two pulses do not overlap temporally until they impinge on the analyzer crystal, where they are spectrally dispersed and temporally stretched to ~ 22 fs (~8 fs FWHM) corresponding to the ~0.24 eV FWHM Si (220) resolution (step 3). The two signals then create the frequency interference with the fringe spacings that are inversely proportional to their time delays (step 4). (See SM for more details.)

**Figure 2 here**

It has been shown that at the onset of amplification ASE and seeded stimulated emission spectra can exhibit gain narrowing (*17, 18, 20*) . Once the superfluorescence takes over, transform limited pulses build up and the emission spectrum features a nearly constant spectral width as the amplification increases (*18*) before spectral broadening and the potential build-up of damped, spectral secondary maxima sets in when approaching saturation (*17-19, 23*) . Further increasing the pump power and/or optical density of the sample can lead to additional broadening and inhomogeneities of the spectral features as well as spatial structures. Figure 3 (A, B) shows examples of broad and inhomogeneous superfluorescence emission spectra from a concentrated solid MnO sample. Also shown are spatial cuts of the spectra obtained for emission along the center direction of the pulse indicated by the white line. We observe similar signals for $MnSO_4$, $Mn_2O_3$, $MnO_2$ and Mn metal foil samples in both superfluorescence and seeded stimulated emission (see SM). Many of the spatial emission profiles show inhomogeneities and some show additional speckle-like features (see Fig. 2B or Fig. S2 in SM). While the origin of these features is not yet fully understood, we note that one possible explanation could be the amplification of multiple field modes starting from noise (*21*). A better understanding and description of the angular and spectral inhomogeneities requires a 3-dimensional numerical simulation and is currently being investigated.

**Figure 3 here**

Strikingly, some of the spectra exhibit regularly spaced fringe patterns along the wavelength dispersive axis (see spectra in Figs. 3C, 3D). We observe these interference fringes in both superfluorescence and seeded stimulated emission from MnO, $MnSO_4$, $Mn_2O_3$, $MnO_2$ and Mn metal foil samples. Our analysis of several hundred single-shot fringe spectra provides the following findings (see SM for more details): 1) occurrence of fringes is rare; 2) fringes occur in spectra with medium to high emission yields but predominantly for saturated emission; 3) fringes are generally equally spaced with most spacings ranging from 0.8 eV to 1.8 eV; 4) fringes can be present in a limited area of the spectrum or extend throughout the whole region (see SM for a discussion of the distribution and statistics of fringe patterns.) Fig. 4 (A-D) shows a series of fringe patterns for $MnO_2$ superfluorescence with spacings ranging from 0.9 eV to 2.5 eV and up to $3.3 * 10^7$ photons/shot (see SM, Tab. S1).

**Figure 4 here**

In the following we show with simulations how the observed fringes are related to the temporal structure of the SASE pump pulse. (See SM for details and a discussion of how we exclude other possible causes.) Fourier analysis of the observed fringes suggests two signals separated by a time $\Delta t$ related to the fringe spacings $\Delta E$ via the Planck constant $\Delta t \Delta E = h = 4.136$ fs-eV. The values for $\Delta E$ (0.9 - 2.5 eV) and corresponding values for $\Delta t$ (1.7 - 4.6 fs) are shown in Fig. 4. We use the 1-dimensional (1D) semi-classical Maxwell-Bloch theory (*23*) to simulate the spectra (see SM for more details). While we do not expect this simulation to reproduce the spectral profile in the saturation region, it provides emission yields for the experimental parameters and links the temporal structure of the SASE pump pulses to the observed fringes. Fig. 4 E-G shows calculated emission spectra for three different pump pulses (shown in the respective insets), and its evolution as a function of relative propagation distance through the gain medium. First, we approximate the SASE pump pulse by a pair of 0.5 fs FWHM Gaussian pulses with a 4 fs spacing, where the relative phase of these pulses is arbitrary (Fig. 4E). The simulated spectrum exhibits well resolved fringes with ~1.0 eV spacing, showing that two spikes in the temporal profile of the pump pulse can lead to X-ray fringes in the superfluorescence. Next, we use realistic temporal profiles for the SASE pump pulses (*10*) – Gaussian noise – having fluctuating spectral and temporal field profiles related by the Fourier transformation (Fig. 4F). Corresponding spectral profiles have been measured for hard X-ray SASE pulses (*25*). The evolution of the emission spectrum as a function of propagation distance for such a typical SASE pump pulse and its emission spectrum at 60% relative propagation is shown in Fig. 4E. While there is some structure in the emitted spectrum, no clear fringe pattern arises (see SM for detailed simulation settings of the pulse parameters). Finally, we use a realistic SASE pump pulse with two dominant temporal spikes separated by ~3.2 fs (Fig. 4G). In this case, interference fringes with constant spacing of ~1.25 eV arise in the superfluorescence. The fact that superfluorescence is a highly nonlinear phenomenon explains why in the weaker spikes of the SASE spectrum might not lower the observed contrast of the interference fringes: A temporal spike only creates superfluorescent emission once it reaches the threshold required for sufficient population inversion. This also explains why fringes are rare. Regarding the likelihood of SASE pulses with predominantly two dominant spikes separated by less than 5 fs, we note that SACLA was running in strong bunch compression mode with <8 fs pulse length. We speculate that the varying occurrence rate of fringes for different SACLA runs reflects variations in the strong electron-bunch compression in the accelerator, causing different temporal profiles of the SASE pulses. Experiments and simulations using various strong bunch compression schemes at LCLS, while different from those at SACLA, have shown that strongly compressed SASE pulses can have so-called 'horn-like' temporal structures that favor the likelihood of having two strong temporal spikes. We further speculate that observed fringe spectra with varying spacings correspond to SASE pulses with more than two strong spikes and modulations resulting from the other spectral inhomogeneities.

**Conclusions**

We have experimental evidence that X-ray superfluorescence and seeded stimulated emission generated by an XFEL SASE pulse with coherence times in the sub-fs range can result in phase-stable fs X-ray pulse pairs. The resulting spectral fringe pattern contains the information about the temporal profile of the interfering X-ray pulses, temporal coherence, delay, and relative phase. The spectral fringe separation directly encodes the time delay of the pulses, which can be determined with sub-fs precision. Our simulations show that with the current experimental resolution we should be able to measure relative fringe spacing differences of ~$5\times10^{-3}$, translating to a ~20 attosecond precision for measuring delays of ~4fs (see SM). Employing an analyzer with higher order Bragg reflection can further improve this precision and enable the discrimination of smaller fringe spacings corresponding to longer pulse delay times. The temporal intensity profile and the relative phase of the pulse pairs that remain undetermined in the current analysis, can potentially be recovered by the application of reconstruction algorithms. To develop phase-sensitive nonlinear X-ray techniques, a stabilization of the relative phase of the two pulses would be required. One potential mechanism is when the temporal ringing of a superfluorescent emission seeds a second burst of superfluorescence, thereby imprinting the phase of the first pulse onto the second (see SM). A future, more robust method for obtaining phase locked pairs of fs x-ray pulses by superfluorescence could be the application of seed pulses at the emission frequency, that have been previously monochromatized and thus having a temporal coherence substantially larger than the XFEL pump-pulse duration. These extensions are currently being investigated in our theoretical studies and will enable coherent nonlinear spectroscopies and nonlinear imaging in the X-ray domain (see e.g. (*5, 8*)) to study vibronic wavepackets at unprecedented spatial and temporal resolution. Further improvement in the generation and control of double superfluorescent pulse pairs will benefit from emerging XFEL pulse shaping techniques (*12, 25-30*). Extending the approach beyond pulse pairs might pave the way to realizing frequency combs in the hard X-ray region.

**Acknowledgements**

The authors would like to thank Joachim Stöhr, Robert Byer, James Cryan, Toru Hara, Shimon Kolkowitz and Philippe Wernet for discussions and Terry Anderson for help with the figures.

**Funding**

The experiment at SACLA was performed with the approval of the Japan Synchrotron Radiation Research Institute (proposal No. 2017B8066). JPSJ KAKENHI is acknowledged for Grant No. 19K20604 (II).

Part of this work was supported by the Department of Energy, Laboratory Directed Research and Development program at SLAC National Accelerator Laboratory, under Contract No. DE-AC02-76SF00515 (UB, YZ).

SSRL Structural Molecular Biology Program is supported by the DOE Office of Biological and Environmental Research, and by the National Institutes of Health, National Institute of General Medical Sciences (including P41GM103393). The contents of this publication are solely the responsibility of the authors and do not necessarily represent the official views of NIGMS or NIH (TK).

Part of this work was supported by the Director, Office of Science, Office of Basic Energy Sciences (OBES), Division of Chemical Sciences, Geosciences, and Biosciences (CSGB) of the Department of Energy (DOE) Contract No. DE-AC02-05CH11231 (JY2, VKY)

DE-AC02-76F00515 (CP), the National Institutes of Health (NIH) Grants No. GM055302 (VKY)

No. GM110501 (JY2), No. GM126289 (JK), and the Ruth L. Kirschstein National Research Service Award (F32GM116423, FDF).

**Author contributions:**

    X-ray experiments: TK, CW, YM, FDF, RAM, DS, MY, II, YI, TO, JY1, VKY, HY, UB

    Sample synthesis: GB, DS, FNS, PMA

    Theory and data analysis: YZ, AH, CP, AB, CW, NR

    Discussion of results: All authors

    Writing – original draft: YZ, AB, NR, UB

    Writing – review & editing: All authors

**Competing Interests**
Authors declare that they have no competing interests

**Data and materials availability:**
All data are available in the main text or the supplementary materials.

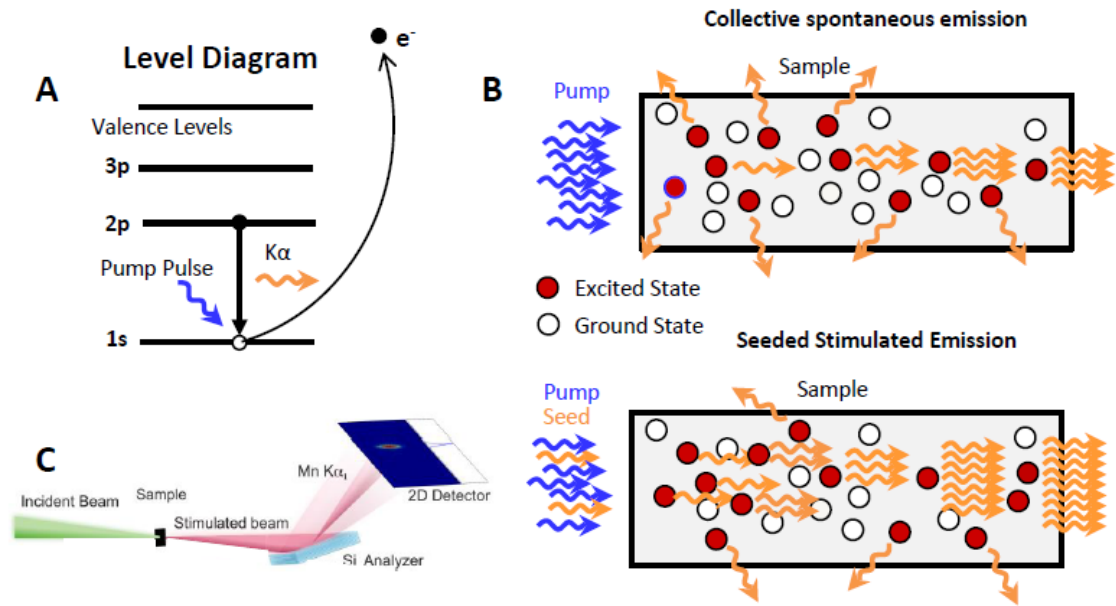

**Fig. 1. Concept of inner-shell X-ray lasing and experimental setup.** (**A**): Level diagram for Kα x-ray fluorescence (orange) following 1s core hole ionization by an incident photon (purple). (**B**): Concepts of the two types of stimulated x-ray emission. The pump pulse (purple) creates 1s core-hole excited states (red). In collective spontaneous emission (ASE and superfluorescence) a spontaneously emitted Kα photon creates amplification by stimulating the emission of a second Kα photon along the direction of 1s core hole-excited states. In seeded stimulated emission, the seed pulse photons (orange) stimulate the emission of Kα photons from 1s core hole excited states along the seeding direction. (**C**): Schematics of the experimental setup.

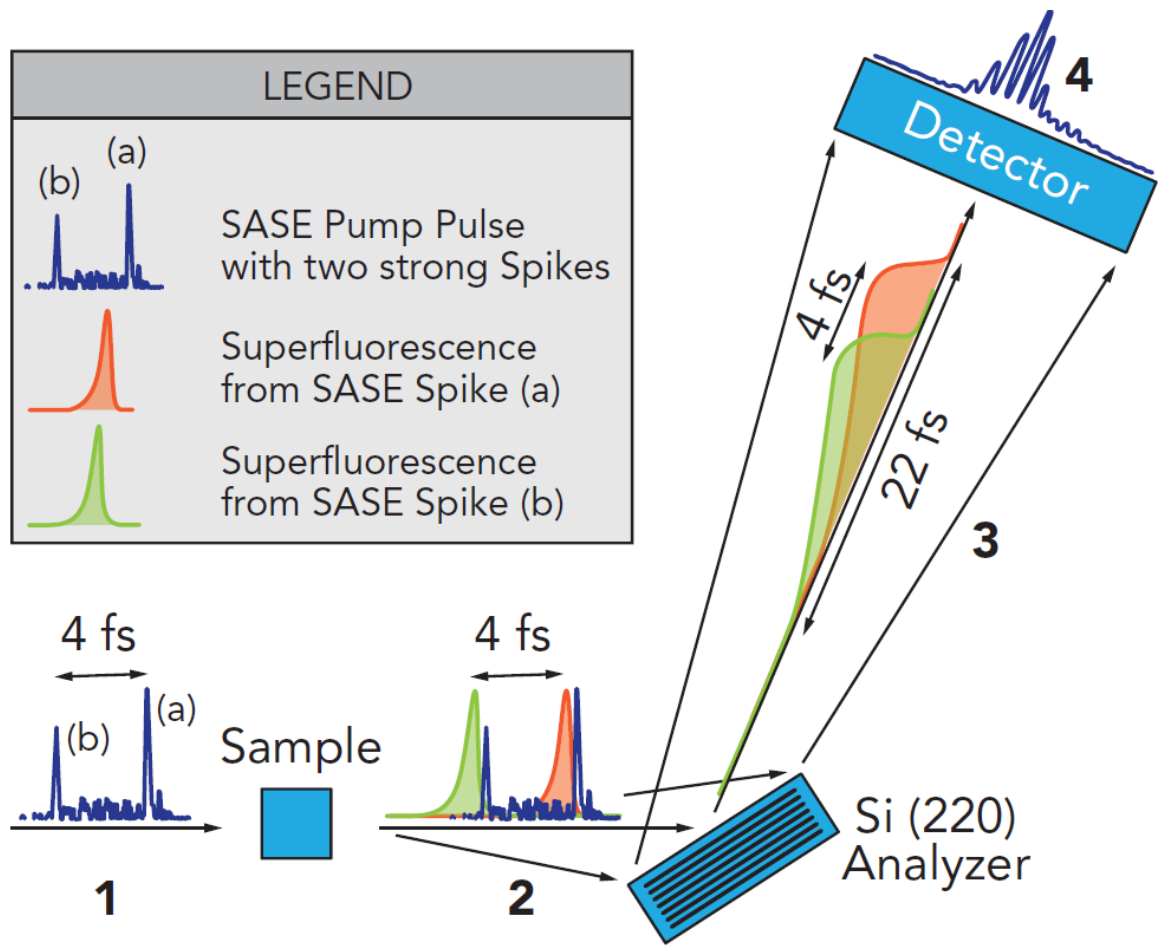

**Fig. 2. Schematics of superfluorescence interference.** In step **1** a SASE pump pulse with two strong spikes (a) and (b) separated by ~4 fs impinges on the sample creating two subsequent superfluorescence pulses. In step **2** the transmitted SASE pulse and the two coherent superfluorescence pulses leave the sample and impinge on the analyzer. The Si (220) analyzer is set at the Bragg angle range corresponding to the Kα spectrum. It rejects the SASE pump pulse and stretches the superfluorescence pulses to ~22 fs duration, corresponding to ~0.24 eV spectral resolution (step **3**). The two stretched pulses create frequency interference along the different Bragg angles that define the dispersive axis of the detector (step **4**).

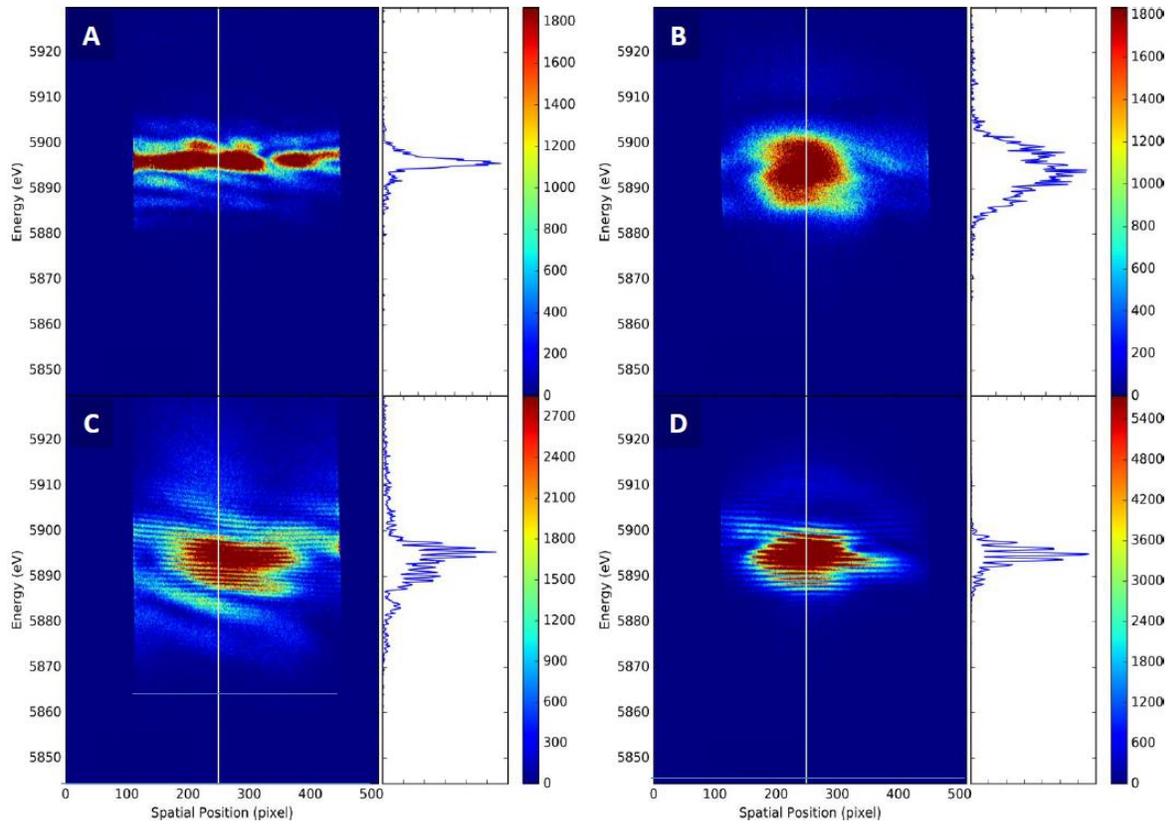

**Fig. 3. Selected single-shot Kα stimulated X-ray emission spectra.** The 2D spectra are shown on the left of each panel, where the vertical axes show the photon energy and the horizontal axes represent the spatial positions on the detector with each pixel corresponding to 50 μm size and ~15.4 μrad angular deviation from the forward direction. The 1D spectra along the cuts (white vertical lines) at the spatial position 250 (pixel value) on the 2D spectral plane MnO (**A**, **B**, **C** without seed pulse, **D** with seed pulse). In each column the 2D s are shown on the right. Superfluorescence spectra with a narrow dominant peak and no obvious fringes (**A**); a broad peak and no obvious fringes (**B**) broad peak and fringes (**C**). Seeded stimulated emission spectrum with fringes (**D**).

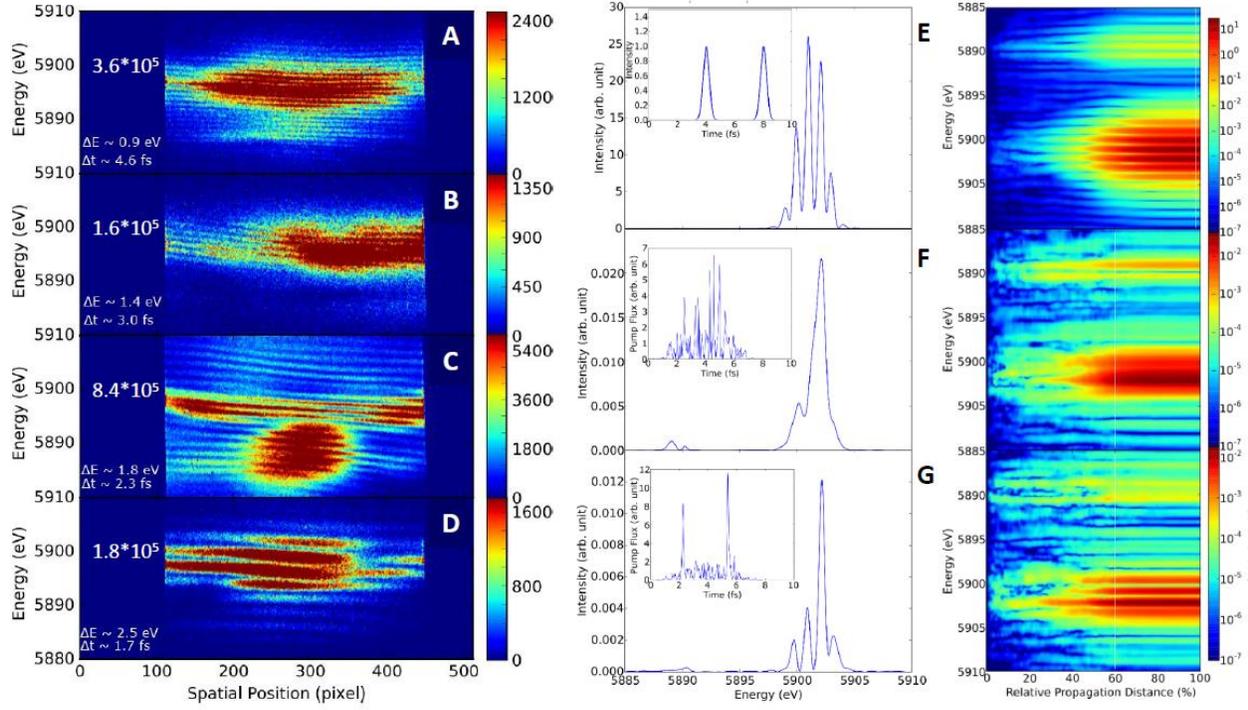

**Fig. 4. Comparison of observed fringes with theory.** Left: Selected MnO$_2$ superfluorescence spectra with increasing fringe spacings $\Delta E$ from ~0.9 to 2.5 eV (**A**-**D**). Number of detected photons are provided for each spectrum. The corresponding time delays $\Delta t$ between two pulses that would cause such interference are shown in the figure. Center and Right: Calculation of interference fringes using 1D Maxwell-Bloch model simulation with a pump pulse consisting of two equal-intensity Gaussians with 0.5 fs FWHM and 4 fs time delay (**E**); a random SASE pump pulse (**F**); and a realistic SASE pump pulse, which has two dominant temporal spikes separated by ~ 3.1 fs (g) (see insets in center panel). The evolution of the emission spectra with relative propagation distance of the pump pulse through the sample (%) is shown to the right. Emission spectra along the white cut lines in (**E**) (98% of the total propagation distance), (**F**) (60% of the total propagation distance), and (**G**) (60% of the total propagation distance) are shown in the center panel.

# Supplementary Materials for

## Generation of Intense Phase-Stable Femtosecond Hard X-ray Pulse Pairs

Yu Zhang[1,*], Thomas Kroll[2], Clemens Weninger[3], Yurina Michine[4], Franklin D. Fuller[3], Diling Zhu[3], Roberto Alonso-Mori[3], Dimosthenis Sokaras[2], Alberto Lutman[5], Aliaksei Halavanau[6], Claudio Pellegrini[6], Andrei Benediktovitch[7], Makina Yabashi[8,9], Ichiro Inoue[8], Yuichi Inubushi[8,9], Taito Osaka[8], Jumpei Yamada[8], Ganguli Babu[10], Devashish Salpekar[10], Farheen N. Sayed[10], Pulickel M. Ajayan[10], Jan Kern[11], Junko Yano[11], Vittal K. Yachandra[11], Hitoki Yoneda[4], Nina Rohringer[7,12,*], Uwe Bergmann[1,13,*]

[1] Stanford PULSE Institute, SLAC National Accelerator Laboratory; Menlo Park, USA

[2] SSRL, SLAC National Accelerator Laboratory; Menlo Park, USA

[3] LCLS, SLAC National Accelerator Laboratory; Menlo Park, USA

[4] The University of Electro-Communications; Chofu, Tokyo, Japan

[5] Linac & FEL division, SLAC National Accelerator Laboratory; Menlo Park, USA

[6] Accelerator Research Division, SLAC National Accelerator Laboratory; Menlo Park, USA

[7] DESY, Center for Free-Electron Laser Science; Hamburg, Germany

[8] RIKEN SPring-8 Center; 1-1-1 Kouto, Sayo, Hyogo 679-5148, Japan

[9] Japan Synchrotron Radiation Research Institute; 1-1-1 Kouto, Sayo, Hyogo 679-5198, Japan

[10] Department of Materials Science and NanoEngineering, Rice University; Houston, USA

[11] Molecular Biophysics and Integrated Bioimaging Division, Lawrence Berkeley National Laboratory Berkeley, USA

[12] Department of Physics, Universität Hamburg; Hamburg, Germany

[13] Department of Physics, University of Wisconsin–Madison; Madison, USA

\* Corresponding author. Email: yu.spacezhang@gmail.com (Y.Z.); nina.rohringer@desy.de (N.R.); ubergmann@wisc.edu (U.B.)

## Materials and Methods

The experiments on were performed at the nanofocus instrument BL3 at the SACLA XFEL (*31-33*). The Mn foil sample was purchased commercially, the Mn(II)O, Mn(II)SO$_4$, Mn(II)Cl$_2$,Mn(II)(acac)$_2$ (acac=acetylacetone), Mn(III)$_2$O$_3$, Mn(III)(acac)$_3$ and Mn(IV)O$_2$ solids were condensed in carbon films at the Ajayan lab at Rice University. The SACLA XFEL was operated in one-color mode for the superfluorescence studies and in two-color mode for the seeded stimulated emission studies. Both modes have been used in previous stimulated emission studies. The two-color seeding mode has been previously used to enhance the K$\alpha_1$ and K$\alpha_2$ lines of Cu foil (*17, 26*). Here, the undulator is split into two sections (*26*), where the first section is tuned to generate the 6.6 keV pump pulse above the Mn K-edge absorption edge, and the second section is adjusted to generate the second color seed pulse tuned to 5.9 keV corresponding to the Mn K$\alpha_1$ emission energy. XFEL pulses are focused through Kirkpatrick-Baez (KB) optics with an estimated 2 mrad vertical divergence and 4 mrad horizonal divergence. The pump pulse energy was ~180 μJ at 6.6 keV photon energy, with a beam size ~200 nm diameter and pulse length of ~8 fs. This corresponds to ~$1\times10^{20}$ W/cm$^2$ pump pulse power. The seed pulse energy at 5.9 keV is estimated at ~2 μJ and similar pulse length. To analyze the K$\alpha$ superfluorescence and seeded stimulated emission spectra we employed a Si (220) crystal (0.24 eV energy resolution) in a horizontal scattering geometry followed by a 2D MultiPort Charged Coupled Device (MPCCD) detector (developed at SACLA) with 512×1024 pixels, each with 50×50 μm$^2$ size. The analyzer had a 1º asymmetric cut to reduce the background from the specular reflection and was previously used in other stimulated emission experiments (*18, 20*). Our numerical simulations were performed with a 1-dimensional (1D) semi-classical Maxwell-Bloch theory (*16, 23, 34*) for the simulation of ASE/superfluorescence. The simulation adopts the following approximations: 1) atomic multiplet theory (*35*) to describe the solid-state electronic structure of the sample; 2) neglecting the effect of 3d spectator holes, which may broaden the emission spectrum; 3) assuming a uniform density of Mn atoms; and 4) neglecting any lateral spatial distribution of the pump pulse and emission process. (See the section "Theory of Maxwell-Bloch simulation" for more details.)

## Supplementary Text

### Examples of fringe patterns in the spectra

As shown in Fig. S1, fringe patterns can be observed in many of the strong S-XES single shots on MnO, MnSO$_4$, Mn$_2$O$_3$ and MnO$_2$ samples with both the seeding and non-seeding experimental modes. Not surprisingly, experiments using a pure Mn metal foil in the seeding mode produced fringes even more easily than the experiments with the various the Mn oxide samples. We believe that the Mn metal foil experiments favor the occurrence of fringe patterns because they exhibit a higher S-XES occurrence due to the higher Mn concentration and absence of other absorbers. In the S-XES experiments on MnCl$_2$, Mn(acac)$_2$ and Mn(acac)$_3$ we observed no strong S-XES signal and no fringes. This might be due to the strong X-ray absorption of the Cl and other atoms in these samples.

As discussed in the main paper, the occurrence of spectral fringes is typically accompanied by the occurrence of other spatial and spectral inhomogeneities (see e.g., Fig. S1(b)). We have previously observed similar inhomogeneities on stimulated X-ray emission experiments with concentrated Mn solutions in the highly saturated S-XES regime. We speculate that these spatial patterns could be due to the breakup of the S-XES beam as the result of multiple filamentations during the propagation of the pump laser pulse, or other phenomena related to the strong lasing regime (*36*). In this study, we focus on the analysis of the spectral fringes, and leave the full analysis of these spatial patterns for the future.

To obtain a more quantitative analysis of the fringes, we provide 1D spectra along at representative spatial positions on the 2D spectral planes. Such spectra taken at pixel position 250 are shown in the respective right panels of the 2D plots in Fig. S1. Another possible way to obtain 1D fringe spectra from the 2D images is to integrate along the spatial axis. However, this integration smears out the fringe patterns as most fringes are not be exactly perpendicular to the spatial direction and slightly curved (see Fig. S1). The origin of these distortions is related to the slight misalignments between S-XES source point, the Si crystal analyzer, and the detector. We have observed similar distortions in a more recent experiment, when we calibrated the analyzer crystal with a monochromatic seed pulse. We therefore only use cuts for our fringe analysis.

Fig. S2 shows two examples of strong single-shot spectra, exemplifying the difference between fringes and other spectral and spatial inhomogeneities. While the spectrum on the left shows fringes as well as other inhomogeneities, the one on the right shows strong inhomogeneities but no fringes. Such a spectrum would not be included on our analysis of the fringe occurrence.

Recognition of fringes

As only a small fraction of the single-shot spectra exhibits fringe patterns, we have to extract spectra with fringes out of a very large amount of raw experimental data (~ 8 TB in total). This is not trivial because there are different types of inhomogeneities (see Fig. S2) and fringe patterns may exist in some spatial and energy regions on the 2D spectral plane but disappear elsewhere. As mentioned above, integrating the signals along the spatial axis smears out the fringe patterns, thus we have to analyze the signals along spatial (vertical) cuts. Multiple choices of cuts add complexity to the data analysis. This is a typical 2D pattern recognition problem in computer science (*37*). Finding an accurate as well as efficient automatic recognition algorithm for identifying all fringe patterns goes beyond the scope of this paper. Here we present a simplified protocol to quickly identify spectra with fringe patterns. For brevity, we only present the main ideas as the following: First, we Fourier transform the 1D signals from selected cuts on the 2D spectral planes and find the peaks on the obtained autocorrelation function curves in the time domain. An ideal fringe pattern will correspond to two autocorrelation peaks well separated in the time domain. However, in reality we usually see multiple 'fuzzy' peaks. We sort the autocorrelation peaks according to intensity and calculate their relative intensity ratios. We apply several empirically found rules that provided us with a good selection the fringe patterns:

1. The ratio of the strongest to the third strongest peak has to be larger than 4/3. We allow one or two strong peaks, and the strongest peak should be sufficiently dominant compared to the third strongest peaks. An ideal fringe pattern only has one strong peak, but we find that the pattern with two strong peaks still may exhibit a fringe pattern.
2. The ratio of the strongest to the fourth strongest peak has to be larger than 2. We find that if the third or even the fourth peaks are relatively strong (violating the condition above or here), the fringe pattern gets blurred and hard to identify.

The parameters in this evaluation model were determined empirically by trial with the goal to exclude non-fringe data as much as possible in the parameter training process. Therefore, we set relatively strict selection criteria and it is likely that we miss some fringe spectra, but we guarantee that more than 90% of our selected spectra have very clear fringe patterns (checked by human eyes). Sampling shows that the remaining 10% of the spectra typically have an ambiguous character with no obvious fringes. We chose this more stringent approach, because it is not our goal to exactly quantify the occurrence of fringes but rather select and characterize spectra, which do exhibit clear fringes. The probabilities discussed in the next section are based on this more stringent approach.

Photons counts and probability of the occurrence of fringes

Given the stochastic nature of the XFEL pump and seed pulses and the highly nonlinear dependence on the exact peak power and overlap (for seeding), the occurrence of fringes varies significantly between runs. We define any shot with S-XES photon counts of $> 10^7$ detector units (energy range defined as 5880.6 – 5899.9 eV in our analysis) as a strong shot. Here, $10^7$ detector units, correspond to $10^7/111 = 9\times10^4$ detected photons per shot. Note that the actual number of emitted photons per shot is a factor 40 higher due to the fact, that the Si crystal Bragg analyzer filters the spectrum dispersively onto the 2D detector (*18*). Some examples for numbers for detector unit and correspond photons per shot (after correction for efficiency) from the shots shown in the main text are provided in Table S1. At this threshold, photons contributing from the seed pulse (when operating in the seeding mode) can be neglected, because the seed pulse is typically two orders of magnitude weaker. Table S1 shows the occurrence probability of strong shots and fringes from different samples with and without seeding in several representative runs.

As illustrated in the Table S2, the occurrence probability for fringes strongly varies from run to run. In most runs, the probability of finding a strong shot is higher than that of finding a shot that exhibits fringes. One of the two exceptions is run 625195 (Mn foil sample with seeding) where we used the most concentrated sample. In that run we observed the highest probability of strong shots (19.1%), and we found that an even higher percentage (27.2%) of all the shots exhibit fringes. (We believe that this specific run could have used a machine tuning that favors the occurrence of two strong SASE spikes.) On the other hand, in run 625668 ($MnO_2$ sample/no seeding) where we observe no strong shots, we also observe no fringes. From the data shown in Table S2 and our overall analysis of the probability of fringe occurrence we find that, generally:

a) Single shots that exhibit fringe patterns are rare and even rarer than strong shots
b) Most of the shots that do not exhibit a strong overall signal do no show fringe patterns
c) Seeding and higher Mn concentration favors strong overall signal and the probability of occurrence of fringes.

Assessing the fringe spacings

Considering the energy resolution of our spectrometer/detector system given by the convolution of the resolution of ~0.24 eV from the Darwin width of the Si (220) Bragg crystal analyzer and that corresponding to the pixel size of the 2D detector (~0.21 eV), we find that the energy spacings between fringes are generally equidistant. We, however, also observed shots with non-equidistant fringe spacings. This is not surprising, given the fact that equally spaced fringes require the interference of exactly two pulses separated in the time domain. The emission spectra have inhomogeneities that can influence the exact location of the fringe maxima. Furthermore, the SASE spectrum of the pump pulse, even with two strong spikes might cause a more complex interference. Fig. S3 shows the case of a very well resolved fringe pattern (using the cut as shown in Fig. S1). To illustrate the precision of the fringe spacings we overlaid a grid with exactly 1.205 eV spacings. All the large peaks align very well, and there might be small deviations at lower intensities from interference of the other spectral inhomogeneities.

Distribution of the average fringe spacings

The distribution of fringe spacings we observed were not related to the sample. A given sample can exhibit a range of different fringe spacings, and different samples can exhibit the same fringe spacing. The former case is illustrated in the four spectra from $MnO_2$ samples shown in Fig. 3 in the main text. To obtain a larger picture of the distribution of fringe spacings for different samples and experimental modes, we calculated the average fringe spacing using the time separation between two dominant peaks in the Fourier-transform of the emission spectra. The distributions of these average fringe spacings are illustrated in Fig. S4. We require that all fringe spacings are greater than 0.2 eV and discard any fringe spacings greater than 4.1 eV, because we cannot reliably identify fringe patterns with such large spacings. The color of the scatters in Fig. S4 represents a fringe quality measure Q, which is defined as the ratio of the intensity of the second strongest and the strongest peaks on the amplitude curve of the Fourier-transformed 1D spectrum. Larger Q usually means cleaner fringe patterns. However, from the scatter plots we could not find visible correlation between emission count, fringe quality, and fringe spacing. We also find that experiments with and without seeding provide statistics that show similar key features. (Note that we have less data points for experiments without seeding.) From the scatter and histogram plots we find that the distributions of fringe spacings are nonuniform in the range of 0.4 – 4.1 eV for all samples and experimental modes. The most dominant distribution ranges are 0.8 – 1.8 eV (corresponding to temporal SASE spike spacings of 5.2 – 2.3 fs) and we find some shots with fringe spacings in the range of 3.0 –3.5 eV (corresponding to temporal SASE spike spacings of 1.38 – 1.18 fs). Distributions in other energy ranges are much less prominent and nearly uniform. We assign the fact that different samples may lead to slightly different fringe distributions, to differences in respective Mn concentration. We speculate that the predominance of fringe spacings that correspond to temporal SASE spikes separated by 2.3 - 5.2 fs, reflects the higher occurrence of such SASE pulses with two strong spikes that have such spacings and the limitations in our experimental setup and conditions. To find signals with smaller fringe spacings (i.e. longer time separation between the SASE spikes) would require a better analyzer resolution and possibly a different bunch compression mode of the SACLA linac. We are considering conducting such studies in the future.

Simulating the fringe contrast

To estimate the degree of phase stability of the two interfering pulses we compare one of our experimentally observed interference spectra with a simulation. The spectrum shown in Fig. S5 (top) is the replotted spectrum from Fig. 2(d). In the first step, we calculate an interference spectrum with 100% contrast, simulating the situation that both pulses are of equal strength and 100% phase stable. We use the product of a Gaussian with 5 eV FWHM and a sine square function (using a peak delay similar to that of the experimental spectrum) to simulate interference fringes with a 1.23 eV fringe spacing, see Fig. S5 (center). In the second step we include the instrumental resolution of our setup by convoluting the spectrum with a resolution function that reflects our experimental setup. We build this function by convoluting a square function with 0.2 eV width to simulate the detector pixel size with a 0.24 eV FWHM Gaussian to simulate the resolution (Darwin width) of the Si (220) analyzer crystal. (We found that using the calculated Darwin curve instead of the Gaussian does not change the result significantly. Either case is an approximation, as we do not know the exact polarization of the seeded emission spectrum.) The result is shown in Fig. S5 (bottom). The very good agreement of the contrasts in the experimental (top) and simulated (bottom) spectra shows that for this shot both interfering stimulated emission pulses were of equal strength and fully phase stable.

Estimating the fringe-space precision

Fig. S6 shows a comparison of two simulated spectra with a 5 eV FWHM Gaussian envelope and interference fringes using cosine square functions with 1 eV and 1.005 eV fringe spacings, respectively. The spectra have been convoluted with the function simulating our experimental resolution (see above). The insets show the shifts of the fourth maxima to each side of the of center with the larger fringe spacings shift to lower (left) and higher energies (right). The combined shift of the two maxima separated by 8 eV is 0.04 eV. Similar size shifts have been observed in XES measurements of ~4 eV-wide manganese Kβ spectra (*38*) using analyzer instrumentation with ~ 0.5 eV resolution (*39*). This comparison and our simulation shows that with our current ~0.3 eV resolution experimental setup it should be possible to discern frequency spacings with $5\times10^{-3}$ relative difference. In the case of an interference with 1 eV fringe spacing caused by two phase-stable pulses separated by ~ 4 fs, this corresponds to a temporal resolution of ~ 20 attoseconds. Analyzer setups with 5-times higher resolution can be achieved experimentally and should further enhance this resolution.

Excluding possible causes for interference fringes

We now discuss possible scenarios of the origin of the fringes and why we can exclude them: 1. Can the fringes be caused by superfluorescence events created from different spectral spikes of the SASE pump pulses? We selected a pump pulse photon energy of 6.6 keV, which is above the Mn K edge where the cross section for 1s core-hole population inversion is essentially constant over the spectral width of the SASE pulse. At this photon energy the emission is non-resonant and independent from any spectral features of the pump pulse. Spectral SASE spikes cannot explain the observed fringes. 2. Can interference of forward and backward emitted superfluorescence signals in the sample generate the temporal delay causing frequency fringe patterns like the Bragg diffraction in a thin crystal (*40*)? This would require 180° diffraction of the backward emitted signal causing a few fs time delay. Both backward superfluorescence and seeded stimulated emission and 180° diffraction in a polycrystalline sample have a much lower probability than forward emission, excluding such a mechanism. 3. Can nonlinear optical phenomena such as self-phase modulation (*36*) or four-wave mixing, known to create spectral interference patterns cause the fringes? Self-phase modulation does not produce equally spaced fringes (*41*) and our 1D Maxwell-Bloch simulations show that much higher pump power than used in our experiment would be required to cause self-phase modulation and four-wave mixing, excluding this mechanism.

Theory of Maxwell-Bloch simulation

The level model of the stimulated X-ray emission processes is illustrated in Fig. S7. The sample is irradiated with an XFEL pump pulse tuned above the 1s core ionization threshold (∼ 6600 eV) and a core hole/ground state population inversion is created. The Kα emissions can take place spontaneously or be stimulated by other spontaneous emissions or seed photons with the same Kα transition frequencies. As shown in the figure, exponential amplification of the emissions finally may lead to lasing in the Kα energy range. Multiple $2p^{-1}$ core hole final states are considered in our simulation. Other dissipation channels such as the Auger decays are also included in our model. As we can see, superfluorescence and seeding experimental modes do not lead to essentially different fringe patterns, so in this study we focus on simulating superfluorescence spectra.

In our 1-dimensional model, the radiation is propagating along the z axis. The X-ray emission is treated as a classical electric field:

$$E(t) = \frac{1}{2}\left[\mathcal{E}(t)e^{-i(\omega t - kz)} + \mathcal{E}^*(t)e^{i(\omega t - kz)}\right], \quad (1)$$

where $\mathcal{E}(t)$ is the field envelope. In order to reduce the numerical dispersion error in integrating the Maxwell equation over long distances, we transform the time variable t into the retarded time $\tau = t - z/c$ (42), where c is the light speed in vacuum. With this choice of time variable, the emission field is recast as

$$E(\tau, z) = \frac{1}{2}\left[\mathcal{E}(\tau, z)e^{-i\omega\tau} + \mathcal{E}^*(\tau, z)e^{i\omega\tau}\right]. \quad (2)$$

Not like the emission radiation, the pump radiation is described as a photon flux. The pump pulse ionizes the sample and is attenuated as it passes through the sample. The attenuation equation is

$$\frac{\partial J(\tau, z)}{\partial z} = -\rho_0(\tau, z) n (\sigma_{1s} + \sigma_{2p}) J(\tau, z), \quad (3)$$

where $\rho_0(\tau, z)$ is the ground state occupation probability, n is the density of the Mn atoms in the sample, $\sigma_{1s}$ and $\sigma_{2p}$ represent the 1s and 2p core ionization cross sections, respectively. The ground state occupation probability evolves as

$$\frac{\partial \rho_0(\tau, z)}{\partial \tau} = -\rho_0(\tau, z)(\sigma_{1s} + \sigma_{2p}) J(\tau, z). \quad (4)$$

After applying the rotating wave approximation and slowly varying envelope approximation (30), the coupled Maxwell equation of the emission field with the retarded time is

$$\frac{\mathcal{E}(\tau, z)}{\partial z} = i\frac{2\pi\omega}{c} P(\tau, z), \quad (5)$$

where the polarization P ($\tau$, z) is determined through

$$P(\tau, z) = 2n \sum_i \mu_i \rho^i_{1s2p}(\tau, z) e^{i(\omega - \omega_i)\tau}, \quad (6)$$

where $\mu_i$ denotes the transition dipole between the $1s^{-1}$ and the i-th $2p^{-1}$ Kα emission final state; $\rho^i_{1s2p}(\tau, z)$ is the off-diagonal reduced density matrix element between the $1s^{-1}$ and the i-th $2p^{-1}$ final state; $\omega_i$ is the emission energy corresponding to the i-th $2p^{-1}$ final state. The Bloch equation governing the motion of the reduced density matrix reads

$$\begin{cases} \dfrac{\partial \rho_{1s1s}(\tau,z)}{\partial \tau} = -(\Gamma_{1s} + \Gamma_t^{\text{rad}})\rho_{1s1s}(\tau,z) + (\sigma_{1s} + \sigma_{2p})J(\tau,z)\rho_0(\tau,z) - \sum_i \mu_i \Im[\mathcal{E}(\tau,z)\rho_{2p1s}^i(\tau,z)], \\ \dfrac{\partial \rho_{2p2p}^i(\tau,z)}{\partial \tau} = -\Gamma_{2p}\rho_{2p2p}^i(\tau,z)\sigma_{2p}J(\tau,z)\rho_0(\tau,z) + \mu_i \Im[\mathcal{E}(\tau,z)\rho_{2p1s}^i(\tau,z)], \\ \dfrac{\partial \rho_{2p1s}^i(\tau,z)}{\partial \tau} = -\dfrac{\Gamma_{1s} + \Gamma_{2p} + \Gamma_i^{\text{rad}}}{2}\rho_{2p1s}^i(\tau,z) + i\left[\rho_{1s1s}(\tau,z) - \rho_{2p2p}^i(\tau,z)\right]\dfrac{\mu_i \mathcal{E}^*(\tau,z)}{2} + S_i. \end{cases} \quad (7)$$

Here $\rho_{1s1s}$ and $\rho_{2p2p}^i$ represent the $1s^{-1}$ and the i-th $2p^{-1}$ core ionized states, respectively. $\Gamma_{1s}$, $\Gamma_{2p}$ denote the 1s and 2p core hole inverse Auger decay lifetimes, respectively. For simplicity we use the same $\Gamma_{2p}$ for different $2p^{-1}$ states. $\Gamma_t^{\text{rad}}$ is the sum of the inverse radiative lifetimes of all studied $1s^{-1} \to 2p^{-1}$ transitions, and $\Gamma_i^{\text{rad}}$ is the inverse radiative lifetime of the i-th $2p^{-1}$ states. $\Im[...]$ means the imaginary part of a complex number. $S_i$ is a random term added phenomenologically to model spontaneous emissions, which reflect vacuum fluctuations of quantum fields. Technically, $S_i$ values can be found in a normal distribution

$$N(x \mid 0, F_i) = \dfrac{1}{\sqrt{2\pi F_i}} e^{-\dfrac{x^2}{2F_i}}. \quad (8)$$

The mean value of this distribution is 0 and the standard deviation is $\sqrt{F_i}$, where $F_i$ is determined as

$$F_i(\tau,z) = \dfrac{\hbar c^2 \rho_{1s1s}(\tau,z)\gamma \Omega \Gamma_{1s}^2}{4\pi^2 \omega_i \mu_i}, \quad (9)$$

where $\gamma$ is the Einstein's A coefficient of spontaneous emission and $\Omega$ is the solid angle in the pulse propagation direction (z direction). $\Omega \approx \arctan(d/L)$ in our case, where d is the XFEL beam focus diameter and L is the total length along the z direction in the sample. The detailed derivation of the equations above can be found in Ref. (*43*).

Simulation settings

Our Maxwell-Bloch simulation is with a 1-D model described in the main text. The focus diameter is set to 200 nm, and the pump pulse duration is 8 fs. The bandwidth of the pump pulse is set to 30 eV. The Rayleigh length of the pump pulse is 20 μm. The propagation length is chosen as 200 μm. Not losing generality, we select MnO as an example sample in our simulation. Because our samples are solid clusters condensed in polymer films, of which the densities of the Mn atoms are nonuniform, we estimate the sample density as 15% of the original density of solid MnO (0.15×5.37 g/cm$^3$). The time step is 0.01 fs, and we propagate for 10000 steps. The spatial step is equal to the speed of light times the time step. In order to see strong lasing effect, we set the pulse energy to 150 μJ. The inverse Auger decay lifetimes of the $1s^{-1}$ and $2p^{-1}$ core hole states are set as 0.8 and 0.32 eV, respectively. The 1s and 2p core ionization cross sections are set as 3.5714×10$^{-24}$ and 1.93263×10$^{-25}$ (a. u.), respectively. These numbers are from atomic relativistic calculations (*44*). The inverse Auger decay lifetime of the $1s^{-1}$ and $2p^{-1}$ core hole states are estimated as 0.80 and 10.33 eV, respectively (*45, 46*). The average natural width of K$\alpha_1$ and K$\alpha_2$ lines is 1.49 eV (*46*), which makes the sum of the inverse lifetimes of all radiative transitions ($\Gamma_t^{\text{ad}}$) equal to 1.49-0.33-0.80=0.36 eV. The inverse lifetime of an individual $1s^{-1} \to 2p^{-1}$ transition ($\Gamma_i^{\text{ad}}$) is determined by partitioning $\Gamma_t^{\text{ad}}$ according to the weights of the corresponding transition dipole squares. We selected 12 strongest emission lines from ligand field multiplet calculations (*46*), which include 6 K$\alpha_1$ and 6 K$\alpha_2$ $2p^{-1}$ core hole final states in our model. The corresponding emission energies, transition dipoles and inverse radiative lifetimes are listed in Table S3.

Simplified model of double-pulse X-ray superfluorescence

The following simplified model can explain the observed fringes: Consider a two-level system (corresponding to K$\alpha_1$ transitions) and emission in the ASE regime (that is, assume that the amount of emitted photons is too low to cause influence on the population inversion. In other words, we assume that the population inversion is determined solely by the pump and incoherent decay). Neglecting the field propagation effects, the field is proportional to the coherence (off-diagonal elements) of the density matrix:

$$E_{(+)} \sim \rho_{eg}, E_{(-)} \sim \rho_{ge},$$

here $E_{\pm}$ denote positive and negative frequency components of the field. The equation for the evolution of the coherence takes the form

$$\frac{d\rho_{eg}(t)}{dt} = -\frac{\gamma}{2}\rho_{eg}(t) + \Gamma_{sp}N_{eff}[\rho_{ee}(t) - \rho_{gg}(t)]\rho_{eg}(t), \tag{10}$$

here $\gamma$ is decoherence rate, $\Gamma_{sp}$ is spontaneous emission rate, $N_{eff}$ is effective number of atoms involved in collective radiation:

$$N_{eff} = \frac{3}{8\pi}n\lambda^2 L,$$

$n$ is the concentration of atoms, $\lambda$ is the wavelength of the fluorescent radiation, $L$ is the effective length of the system.

Assuming that the decay of the core-excited states is much shorter than the pump pulse duration (which is not crucial for considerations below, but simplifies the resulting expressions), the population inversion can be represented as

$$\rho_{ee}(t) - \rho_{gg}(t) \simeq \frac{\sigma J(t)\rho_0(t)}{\Gamma_K}(1 - \frac{\Gamma_{sp}}{\Gamma_L})$$

here $J$ is the pump photon flux, $\sigma$ is cross-section for ionization from the K shell, $\rho_0$ is an occupation of the ground (neutral) state, $\Gamma_{K,L}$ are inverse lifetimes of the holes at the K and L shells.

The solution of Eq. (10) is

$$\rho_{eg}(t) = \rho_{eg}(0)Exp\left(-\frac{\gamma}{2}t + \int_0^t k\rho_0(t')J(t')dt'\right), \quad k = \frac{\Gamma_{sp}N_{eff}}{\Gamma_K}(1 - \frac{\Gamma_{sp}}{\Gamma_L}),$$

here $\rho_{eg}(0)$ stems from quantum fluctuations, it will be discussed below.

Consider a pump consisting of two pulses. For simplicity, consider two delta-function-like pulses arriving at times $t_1$ and $t_2$:

$$J(t) = f_1\delta(t - t_1) + f_2\delta(t - t_2).$$

For such pump, coherence takes the form

$$\rho_{eg}(t) = \rho_{eg}(0)Exp\left(-\frac{\gamma}{2}t + F_1\Theta(t - t_1) + F_2\Theta(t - t_2)\right), \quad F_1 = k\rho_0(t_1)f_1, \quad F_2 = k\rho_0(t_2)f_2, \tag{11}$$

here $\Theta(t)$ is the Heaviside step-function. The initial value of the coherence $\rho_{eg}(0)$ – which triggers the ASE process – is conditioned by the seed pulse or by quantum fluctuations constituting spontaneous emission, whichever is larger. Further, consider the case of no seed pulse. When the amplification process starts, the quantum fluctuations result in random values of $\rho_{eg}(t_1)$. But for a particular measurement (particular shot), some realization of it, let us name it $\delta\rho_{eg}^{(1)}$, is being picked up by the ASE process and amplified by a factor of $e^{F_1}$. More rigorous reasoning can be found in Ref. (28), where a strict proof was given for a case of 0D two-level system ensemble possessing initial population inversion. After the first pump pulse, the created coherence exponentially decays. Assuming $\gamma(t_2 - t_1)$ to be at least ~1, by the moment $t_2$ there are two options: $F_1 > \gamma(t_2 - t_1)$ -- then, the decayed initial coherence is still larger than the fluctuations and can be used a seed; or $F_1 < \gamma(t_2 - t_1)$ and the fluctuations are stronger than the decayed coherence from the first pulse. As result, Eq. (11) can be rewritten as

$$\rho_{eg}(t) = \delta\rho_{eg}^{(1)} e^{F_1} e^{-\frac{\gamma}{2}(t-t_1)} \Theta(t-t_1)\Theta(t_2-t) + \delta\rho_{eg}^{(2)} e^{F_2} e^{-\frac{\gamma}{2}(t-t_2)} \Theta(t-t_2), \tag{12}$$

here $\delta\rho_{eg}^{(2)}$ is given by $\delta\rho_{eg}^{(1)} e^{F_1} e^{-\frac{\gamma}{2}(t_2-t_1)}$ (decayed coherence from the first pulse) if it is stronger than quantum fluctuations, or $\delta\rho_{eg}^{(2)}$ is another realisation of quantum fluctuation (in this case it has, in general, the absolute value and the phase different from $\delta\rho_{eg}^{(1)}$).

The resulting spectral decomposition of the ASE field is obtained by Fourier transform, from Eq. (12) we arrive at

$$E_{(+)}(\omega) \sim \frac{1}{i\omega + \gamma/2} (\delta\rho_{eg}^{(1)} e^{F_1} e^{-i\omega t_1} + \delta\rho_{eg}^{(2)} e^{F_2} e^{-i\omega t_2}) \tag{13}$$

here we have assumed $\gamma(t_2 - t_1) \gg 1$. The spectrum is given by

$$I(\omega) \sim E_{(+)}(\omega) E_{(-)}(\omega)$$
$$\sim \frac{1}{\omega^2 + (\gamma/2)^2} \left( |\delta\rho_{eg}^{(1)}|^2 e^{2F_1} + |\delta\rho_{eg}^{(2)}|^2 e^{2F_2} + 2Re[\delta\rho_{eg}^{(1)} \delta\rho_{eg}^{(2)*} e^{F_1+F_2} e^{-i\omega(t_1-t_2)}] \right)$$

Whenever the quantities $\delta\rho_{eg}^{(1)} e^{F_1}$ and $\delta\rho_{eg}^{(2)} e^{F_2}$ are comparable in size, the spectrum exhibits fringes with period $\hbar/(t_2 - t_1)$.

## Fringe formation in the temporal domain

In the previous part of the Supplementary Information, the appearance of the spectral fringes was explained based on formal Fourier transformation of electromagnetic field temporal profile consisting of two pulses. Here we will present a deeper discussion on how diffraction on an analyzer crystal produces an interference pattern out of two non-overlapping pulses.

To this end, consider a transformation of the temporal field amplitude under the diffraction from a semi-infinite perfect crystal. The problem of X-ray diffraction from a crystal is solved straightforwardly for each Fourier component of the field – each component is reflected with reflection coefficient being dependent on deviation from Bragg condition:

$$E_d(\theta, \omega) = E_0(\theta, \omega) R\left(\theta + \frac{\omega}{\omega_B} \tan \theta_B\right), \tag{14}$$

here $\theta$ and $\omega$ are deviations from angle $\theta_B$ and X-ray frequency $\omega_B$ for which Bragg conditions are satisfied, $R$ is a reflection coefficient that in our case of symmetric (220) reflection of Si has the form

$$R(x) = y - \text{sign}(Re(y))\sqrt{y^2 - 1}, \quad y = \frac{x \sin 2\theta_B + \chi_0}{\sqrt{\chi_G \chi_{\underline{G}}}}, \tag{15}$$

here $\chi_0$ is a uniform component of crystal X-ray susceptibility, $\chi_G(\chi_{\underline{G}})$ are periodic – corresponding to reciprocal lattice vector $G$ (-$G$) – components of crystal X-ray susceptibility (for a detailed discussion see e.g. Ref. (47). Thanks to the linearity of X-ray diffraction, the temporal profile of the diffracted pulse can be obtained as a superposition of diffracted Fourier components:

$$E_d(\theta, t) = \int d\omega\, e^{-i\omega\tau} E_d(\theta, \omega) = \int dt' E_0(\theta, t') R(\theta, t - t'), \qquad (16)$$

here $E_{0,d}(\theta, t)$ are envelopes of the incoming (diffracted) pulse (the frequency of the carrier plane wave is $\omega_B$), $\tau = t - \vec{n}_d \vec{r}/c$ is retarded time (since we will be not interested in spatial field profile, the dependence on $\vec{r}$ is omitted further for simplicity). The reflection coefficient in the temporal domain $R(\theta, t)$ is given by Fourier transformation of Eq. (15), following the integration procedure described in Appendix A of Ref. (47), one obtains for our case

$$R(\theta, t) = e^{i\omega_d(\theta)t} \frac{J_1(t/T_\Lambda)}{i\, t} e^{-t/T_a} H(t), \qquad (17)$$

$$\omega_d(\theta) = -\omega_B \theta \cot\theta_B,\ T_\Lambda = \frac{2\Lambda}{c} \sin\theta_B = \frac{2\sin^2\theta_B}{\omega_B \sqrt{\chi_G \chi_{\underline{G}}}},\ T_a = \frac{2\sin^2\theta_B}{\omega_B \operatorname{Im}(\chi_0)},$$

here $H(t)$ is Heaviside step-function, $\Lambda$ is extinction-length, here we have assumed Bragg conditions being corrected for refraction effects and hence have taken only imaginary part of $\chi_0$ in the second exponent.

Equations (16) and (17) describe the action of the analyzer crystal in the temporal domain. According to the Bessel-function term and decaying exponent term in Eq. (17), the pulse is stretched to a duration determined by extinction length and absorption length (the contribution of extinction length is prevailing). In addition to stretching, according to the first oscillatory exponent term in Eq. (17), the pulse is convolved with the exponent oscillating at frequency $\omega_d(\theta)$. Assuming the stretching time to be long, this convolution would result in obtaining the Fourier component with frequency $\omega_d(\theta)$ of the incoming pulse $E_0(\theta, t)$. Hence, the detector placed at angle $\theta$ (that counts the intensity integrated over infinite time) would give a signal proportional to the spectral power at frequency $\omega_d(\theta)$ of the incoming pulse $E_0(\theta, t)$ – as expected from the analyzer. The accuracy of extracting the Fourier component is determined by the duration of the stretched pulse – by the extinction-length – as expected.

The Eqs. (16) and (17) can be used to illustrate the formation of the fringe pattern. Namely, consider two pulses separated by a delay time $t_{del.}$, for convenience let us assume that the pulse duration is much shorter than the delay time, and the delay time is much shorter than the stretched pulse duration. In this case, from Eqs. (16) and (17) it follows that $E_d(\theta, t)$ will be given by the sum of two contributions with relative phase $e^{i\omega_d(\theta)t_{del.}}$. Hence, for deviations $\theta$ such that $\omega_d(\theta) t_{del.} = (2n + 1)\pi$ one would observe fringe minima and for $\omega_d(\theta) t_{del.} = 2n\pi$ one would observe fringe maxima ($n$ is an integer number). As expected, the same conclusion follows from the spectral treatment presented in the previous sections.

The considerations given above are illustrated in Fig. S8. The case of two pulses corresponding to the exponential decay of impulsively created polarization is considered:

$$E_0(\theta, t) = e^{-\frac{\gamma}{2}t} H(t) + e^{-\frac{\gamma}{2}(t - t_{del.})} H(t - t_{del.}),$$

here the parameter $\gamma$ corresponds to FWHM of the spontaneous emission profile and is taken as 1.48 eV, delay time is $t_{del.} = 4$ fs, the corresponding temporal profile is shown as orange and green solid lines (plotted 10 times decreased in scale) in Fig. S8 (b), the angular distribution is assumed to be flat within the range of interest. The corresponding spectral profile is straightforwardly obtained as

$$I(\omega) = 2 \frac{1 + \cos(\omega\, t_{del.})}{\omega^2 + (\gamma/2)^2},$$

it is shown as the brown dotted line in Fig. S8 (a). To obtain the registered spectrum, this expression should be convoluted with the resolution function of the analyzer (48)

$$I_d(\omega_d) = \int d\omega' I(\omega_d - \omega') \left| R\left(\frac{\omega'}{\omega_B} \tan \theta_B\right) \right|^2,$$

it is depicted by the brown solid line in Fig. S8 (a), the resolution function is shown by the magenta line (parameters for Si (220) was used, the susceptibilities were taken as an average between σ and π polarization). As expected, the same profile can be obtained based on time-domain calculations. Examples for three detuning angles (corresponding to detuning frequencies according to $\omega_d(\theta)$) are presented in Fig. S8 (b)-(d) where $Im(-E_d(\theta, t))$ is shown. The detected intensity is given by the integral of the field intensity over time, the results of which are marked on Fig. S8 (a) by points in the corresponding color (black for Fig. S8 (b), blue for Fig. S8 (c), red for Fig. S8 (d)). In the case of Fig. S8 (b) – corresponding to exact Bragg conditions (corrected by refraction) – the diffracted pulses have no modulation in time (since $e^{i\omega_d(0)t} = 1$) and are added up in phase. As a result of constructive interference, the detected intensity is larger than in the case of incoherent summation of intensities. In the case of Fig. S8 (c) – corresponding to the first fringe minimum – the modulation in time results in destructive interference between diffracted pulses (since here $e^{i\omega_d(\theta)t_{del.}} = -1$) and decreased detected intensity. If the analyzer would be ideal, then the diffracted pulses would be stretched infinitely, and the contribution of intensity until $t_{del.}$ would be infinitesimally small – leading to the absence of the detected intensity (as for the ideal case denoted by the dotted line on Fig. S8 (a)). In the case of Fig. S8 (d) – corresponding to the side maximum – the modulation in time again leads to constructive interference ($\omega_d(\theta)t_{del.} = 2\pi$) similar to the case of Fig. S8 (b).

Role of the emission angle fluctuation on the detected spectrum

The detected energy spread of the superfluorescence pulses is quite large (about tens of eV). However, the carrier frequency of the superfluorescence is determined by the atomic inner-shell transition frequency, and there is no evident reason for it to fluctuate by tens of eV. Here we will demonstrate that one of the possible explanations of the large observed energy spread is an interplay between emission angle fluctuation and analyzer performance.

Consider a pulse with spectral-angular distribution $I_0(\theta, \omega)$ impinging on the analyzer. Following Eq. (14), the intensity detected at an angle $\theta_d$ is

$$I_d(\omega_d(\theta_d)) = \int d\omega' I_0(\theta_d, \omega_d(\theta_d) - \omega') \left| R\left(\frac{\omega'}{\omega_B} \tan \theta_B\right) \right|^2. \tag{18}$$

In the case of the spectral distribution of incoming pulse much broader than the Darwin width, the registered signal is close to cut through the distribution $I_0(\theta_d, \omega_d(\theta_d))$. If the beam has an angular deviation $\theta_a$ (i.e., its angular distribution has the maximum at an offset $\theta_a$) and its angular distribution is sharper than the frequency distribution, then the registered signal would appear at a position corresponding to frequency $\omega_d$ ($\theta_a$), and would have intensity corresponding to the tail of the frequency distribution, namely $I_0$ ($\theta_a, \omega_d$ ($\theta_a$)). Since the superfluorescent pulses have large intensity fluctuation, a strong pulse with an offset $\theta_a$ and spectrum centered at $\omega_B$ could look on the detector as a weaker pulse with spectrum shifted towards $\omega_d$ ($\theta_a$).

Fig. S9 shows a numerical illustration of this case. Here the same parameters as in Fig. S8 were used, except for the angular profile. In Fig. S8 we assumed it to be flat, in Fig. S9 (a) we assume it to be Gaussian with 0.5 mrad FWHM centered around Bragg position, Fig. S9 (b) – shifted by 0.8 mrad (corresponding to 7 eV according to the dependence $\omega_d$ (θ) presented in Eq. (17)). Fig. S9 (c) shows that – for the beam with angular deviation from Bragg conditions – the maximum of the detected intensity is shifted to a value close to $\omega_d$ ($\theta_a$). The shifted pulse appears to be ~70 times weaker than the unshifted one, however, for a strong superfluorescent pulse it may be observed. This illustration shows

that – for the rigorous data treatment – modeling of full spectral-angular distribution would be necessary. This requires a 3D simulation, which is beyond the scope of the current work and will be considered in future publications.

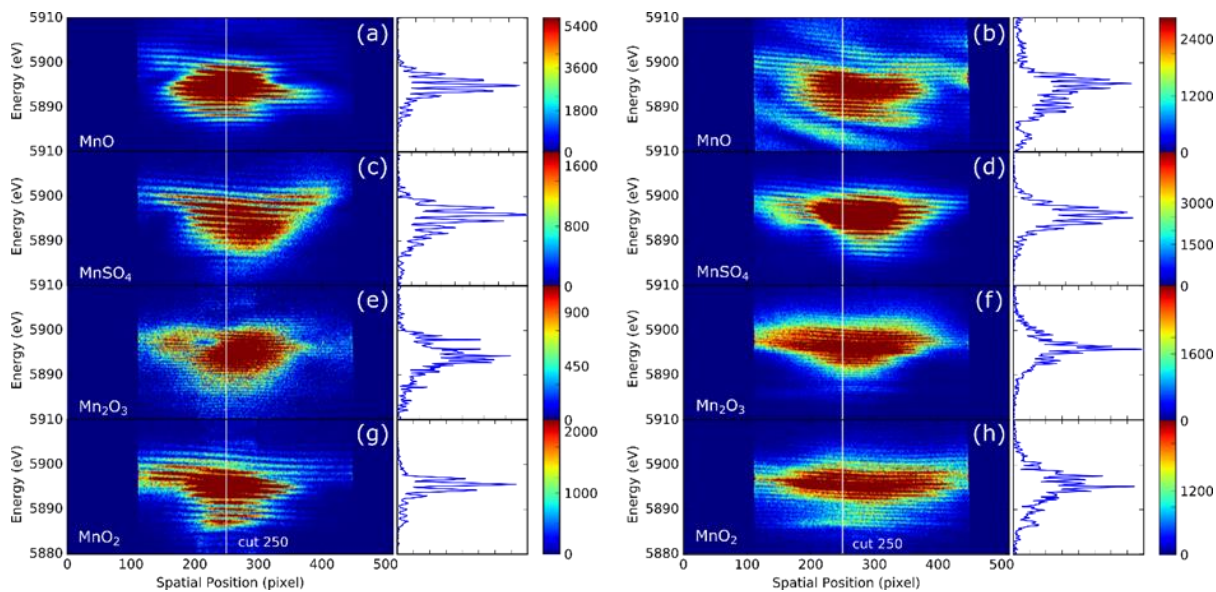

**Fig. S1.**

A series of single-shot Kα stimulated X-ray emission spectra of Mn species with different oxidation states. In each column the 2D spectra are shown on the left where the horizontal axis represents the spatial positions on the detector and the vertical axis denotes the photon energy. The 1D spectra along the cuts at the spatial position 250 (pixel value) on the 2D spectral planes are shown on the right of each column. Spectra in the left column (a, c, e, g) were obtained with seed pulses, spectra in the right column (b, d, g, h) were obtained without seed pulses.

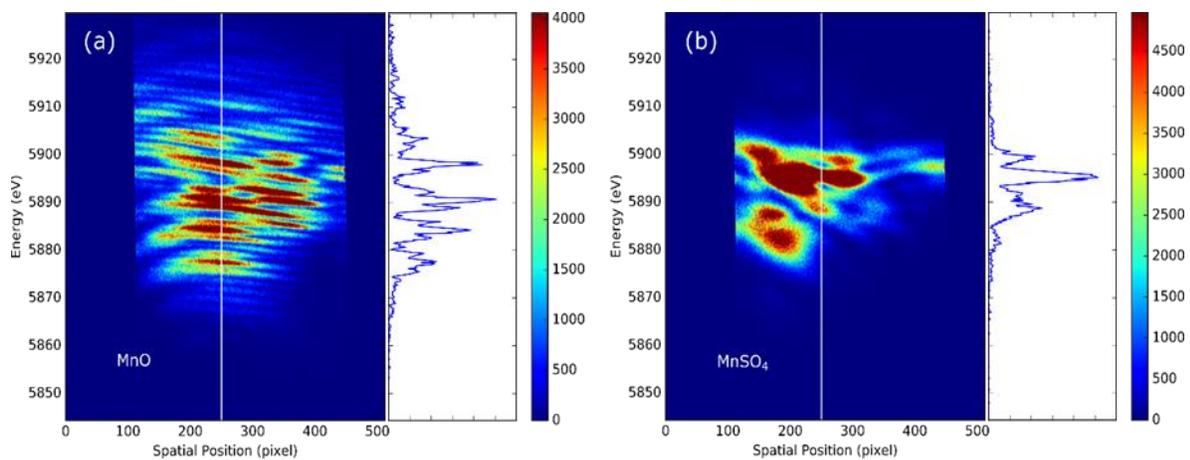

**Fig. S2.**
Single-shot Kα stimulated X-ray emission spectra of MnO and MnSO4. In (a) fringes in a large energy range is shown, and in (b) scattered spatial patterns are presented. The 1D spectra along the cuts at the spatial position 250 (pixel value) on the 2D spectral planes are shown on the right.

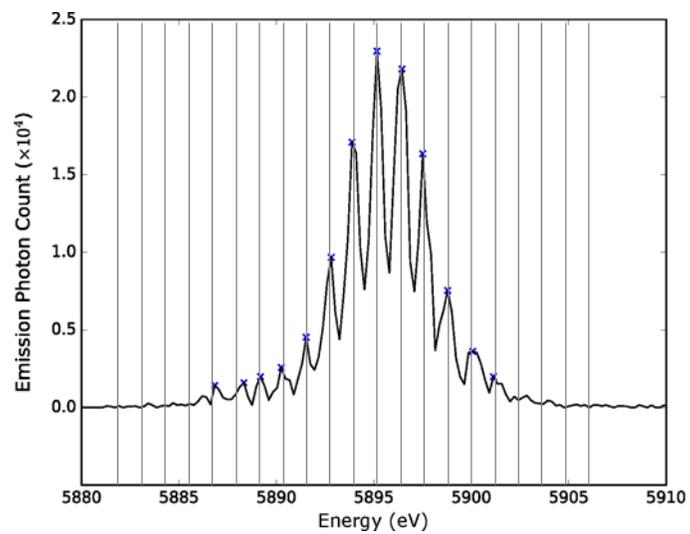

**Fig. S3.**
Superfluorescence spectrum of MnSO$_4$ at the pixel 250 line shown in Fig. S1(d). Spikes are labeled by blue crosses. The energy spacings between the spikes are almost equal, as shown by the grid with 1.205 eV spacings.

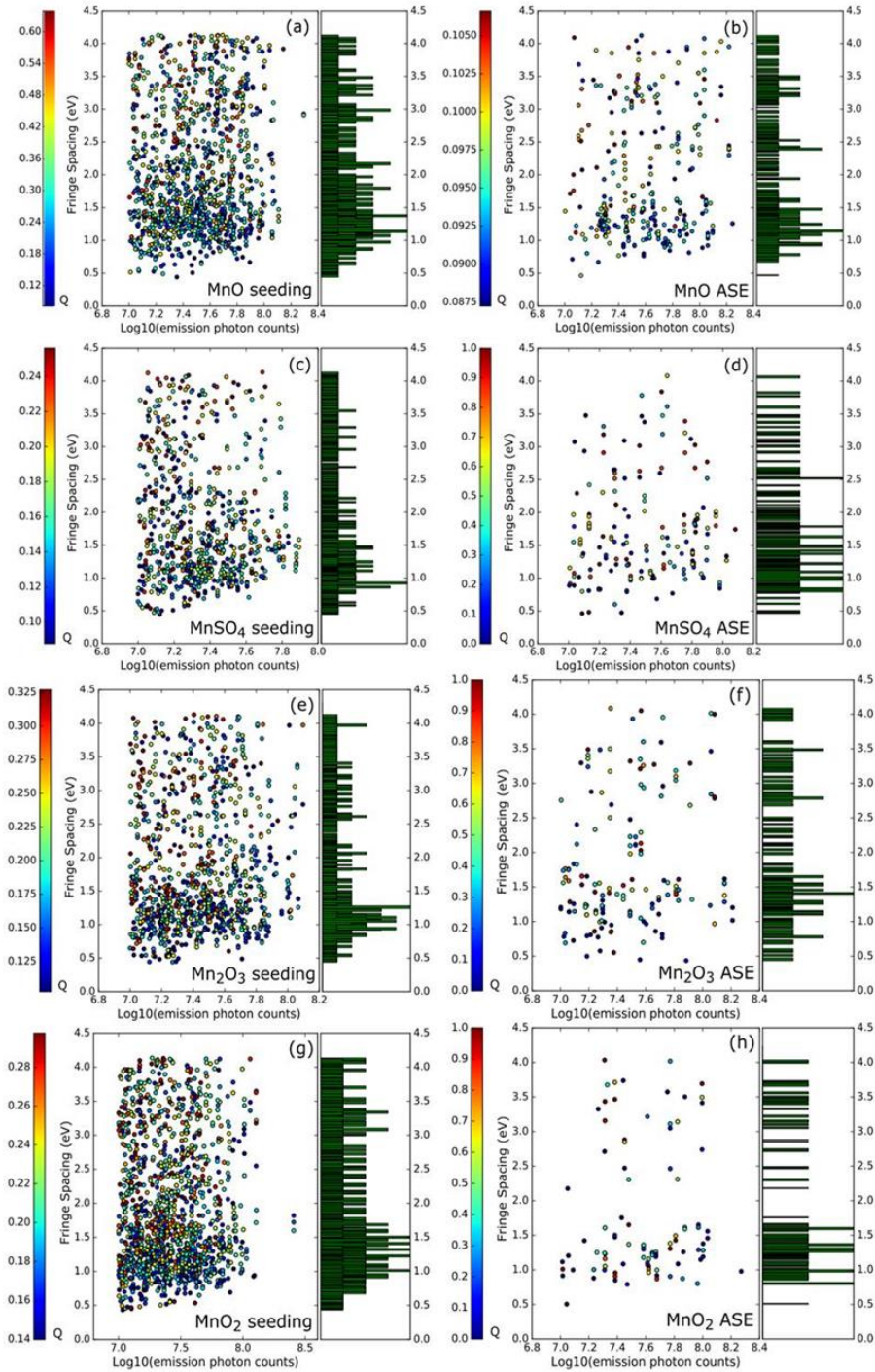

**Fig. S4.**

Distribution of average fringe spacings of different samples and experimental modes. Photon counts are in detector units as described in the main text. The color of scatters represents a crude fringe quality measure Q which is explained in the text.

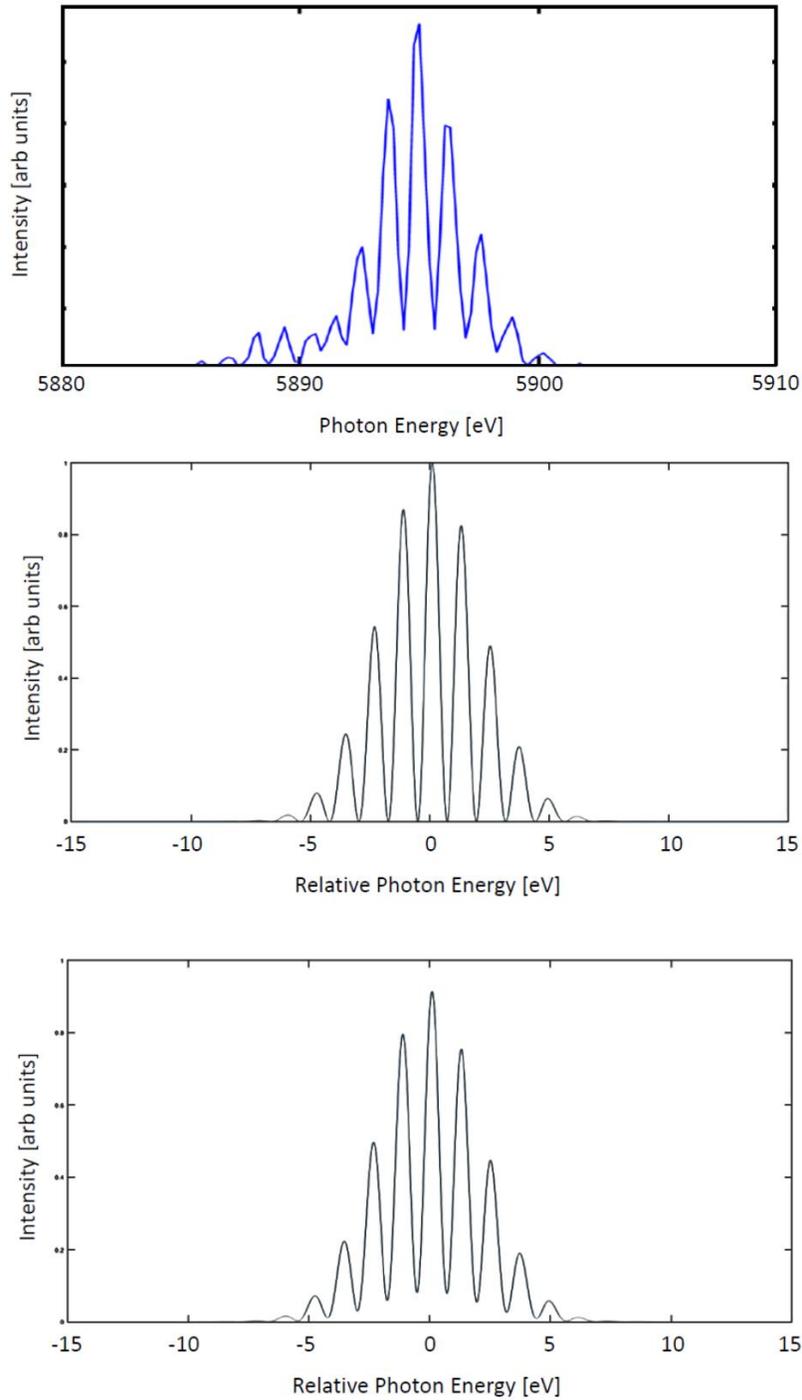

**Fig. S5.**

Experimentally observed seeded stimulated emission spectrum with fringes (top) (This spectrum is also shown in Fig. 2(d)). Simulation of an interference with 100% contrast using the product of a Gaussian with 5 eV FWM and a sine square function with 1.23 eV fringe spacing (center). Convolution of the simulated spectrum shown in the center with a resolution function corresponding to our experimental setup (bottom).

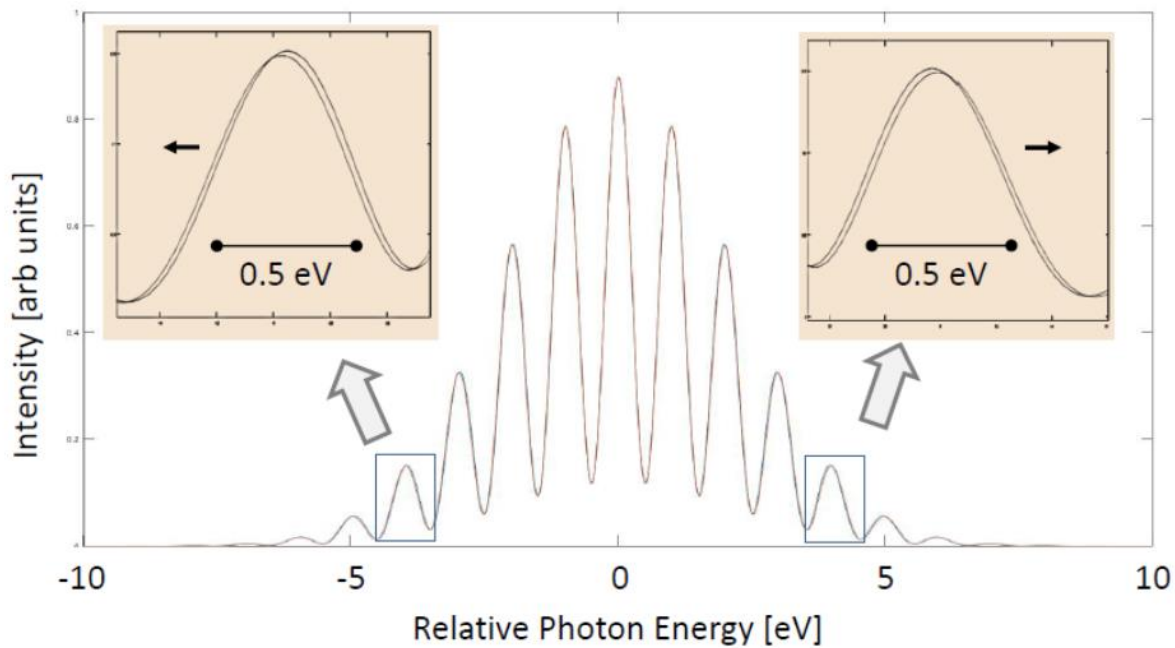

**Fig. S6.**
Simulation of spectra using a 5 eV FWM Gaussian envelope and sine square functions with 1 eV and 1.005 eV fringe spacings, respectively. The spectra are convoluted with the resolution function corresponding to our experimental setup. The insets show two sets of fringe maxima that are 8 eV apart. The combined outward shift of the maxima corresponding to the spectrum with 1.005 eV fringe spacing is 0.04 eV.

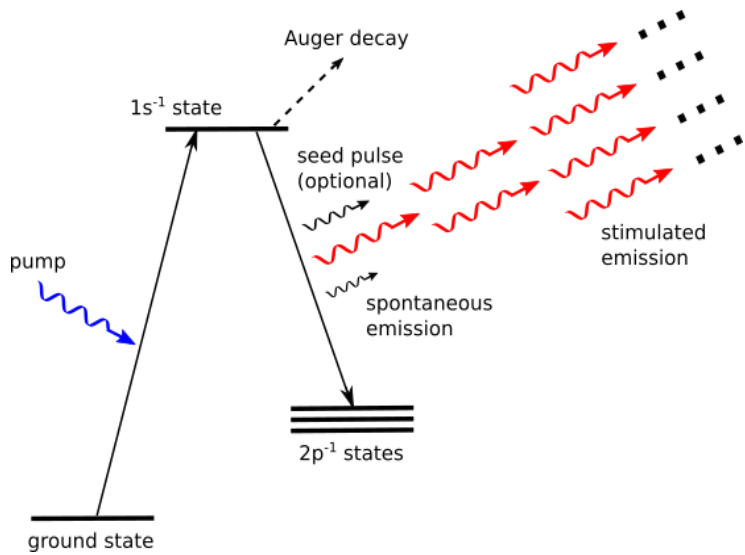

**Fig. S7.**
Level scheme of the modeled stimulated X-ray emission process. The sample is ionized by an XFEL pump pulse and a population inversion of the 1s core hole state is created. The 1s−1→ 2p−1 emission can happen spontaneously or be stimulated by other spontaneous emission photons or a seed pulse tuned at the transition frequency.

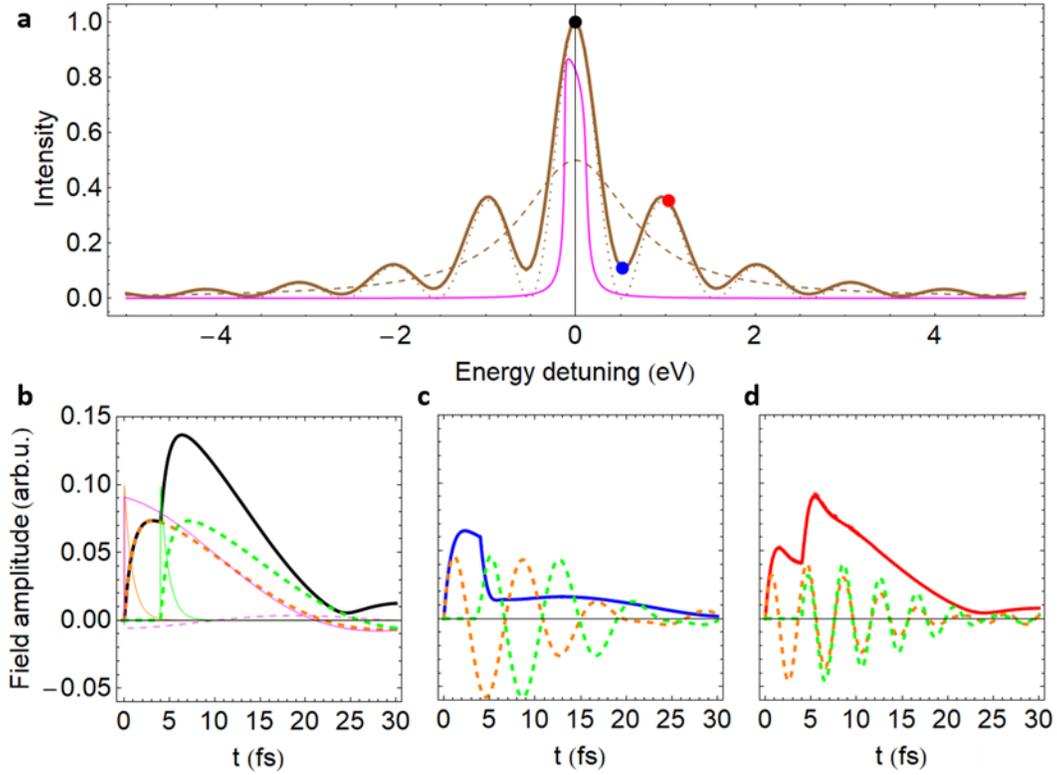

**Fig. S8.**

Illustration of fringe formation in the temporal domain. (a) – calculation of fringe spectral pattern on the detector (solid brown curve, normalized to 1 at maximum), dotted brown line – fringe pattern for an ideal analyzer (normalized to 1 at maximum), dashed brown line – spectrum for incoherent summation of the pulses, solid magenta line – resolution function. (b)-(d) temporal profiles of each of the pulses after the diffraction (orange and green dashed lines) and the total field magnitude ((b) – black, (c) – blue, (d) – red solid lines). The resulting integrated intensity values at corresponding angular positions (at detunings $\omega\_d\,(\theta)$ ) are denoted by points of corresponding colors on (a). (b) and (d) correspond to constructive interference (fringe maxima), (c) – to destructive interference (fringe minimum). In (b) the solid orange and green lines correspond to the temporal profile of incoming pulse (plotted 10 times decreased in scale), dashed and solid magenta lines show real and imaginary parts of R(0,t). See further discussion in the text.

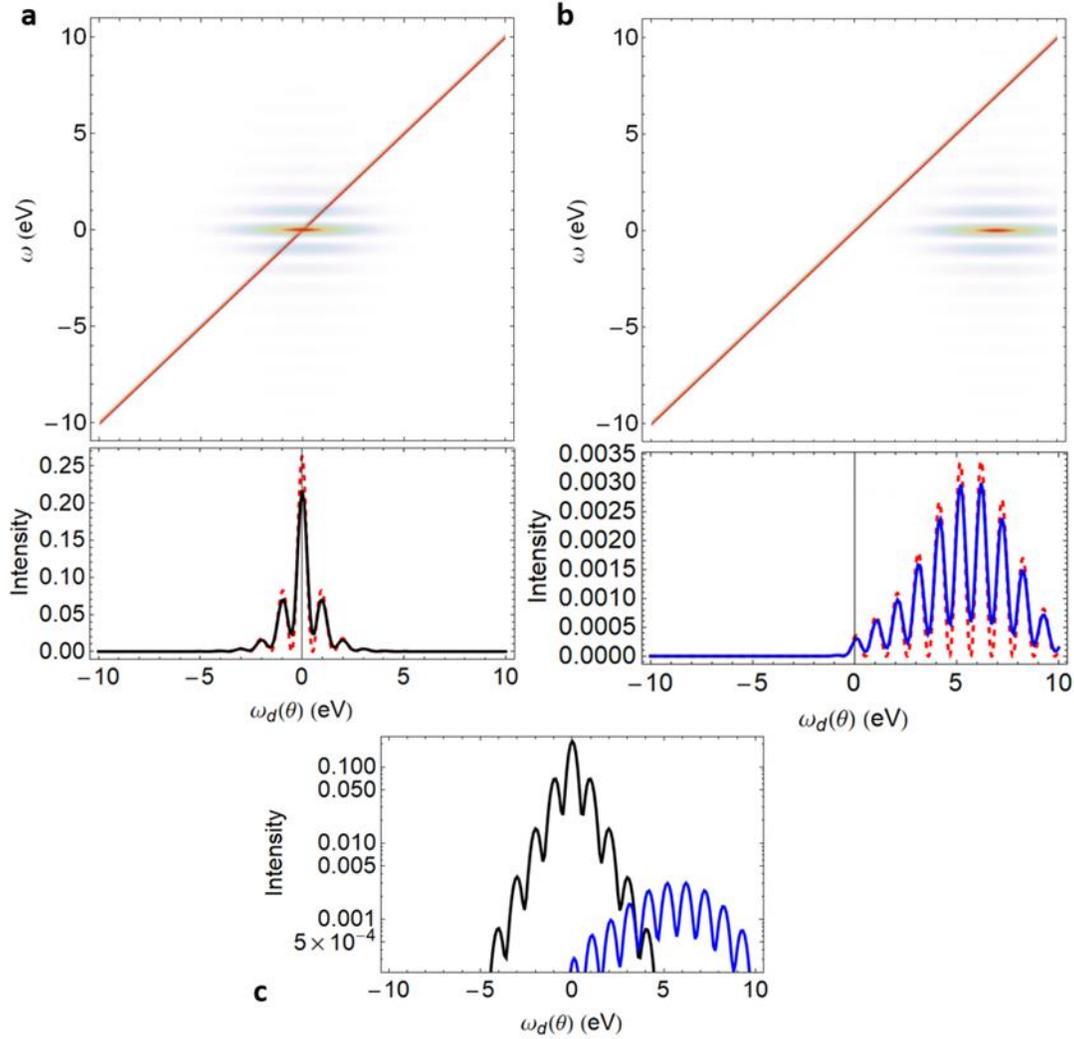

**Fig. S9.**
Illustration of detecting a shifted spectrum from a pulse with angular offset. (a) Spectral-angular distribution $I_0(\theta,\omega)$ of the incoming pulse (the horizontal axis ($\theta$) is represented as $\omega_d(\theta)$) and the resolution function (with the selected scaling appears as a diagonal line). For the spectral part of the profile, the same parameters as in Fig. S8 were used. The angular profile was taken as Gaussian with 0.5 mrad FWHM. The lower panel shows the detected spectral profile (solid black line, calculated according to Eq. (18) ) and a simplified expression for it $I_0(\theta_d,\omega_d(\theta_d))$ (dashed red line). (b) – same as (a), but the angular part of the distribution has a 0.8 mrad offset (corresponding to 7 eV when recalculated by $\omega_d(\theta)$). The detected spectral profile (showed in the lower panel by a solid blue line) has a maximum shifted to ~ 6 eV. (c) – comparison of registered spectral profiles (presented in (a) and (b) ) in logarithmic scale. See further discussion in the text.

| Spectrum | Number of Detector Units per Shot | Number of Detected Photons per Shot | Estimated Number of Photons per Shot |
|---|---|---|---|
| Fig. 3C | $4.258 * 10^7$ | $3.836 * 10^5$ | $1.534 * 10^7$ |
| Fig. 3D | $9.185 * 10^7$ | $8.275 * 10^5$ | $3.310 * 10^7$ |
| Fig. 4A | $4.025 * 10^7$ | $3.626 * 10^5$ | $1.450 * 10^7$ |
| Fig. 4B | $1.804 * 10^7$ | $1.625 * 10^5$ | $6.500 * 10^6$ |
| Fig. 4C | $9.340 * 10^7$ | $8.414 * 10^5$ | $3.366 * 10^7$ |
| Fig. $D | $2.042 * 10^7$ | $1.840 * 10^5$ | $7.359 * 10^6$ |

**Table S1.**

Numbers of photons per selected single shot spectra

| Run No. | Sample/Mode | Probability of strong shots [%] | Probability of shots with fringes [%] | Probability of both strong shots and fringes [%] |
| --- | --- | --- | --- | --- |
| 625131 | MnO/seeding | 5.3 | 4.0 | 2.5 |
| 625153 | MnO/seeding | 7.3 | 4.7 | 4.4 |
| 625195 | Mn/seeding | 19.1 | 27.2 | 18.0 |
| 625295 | $Mn_2O_3$/seeding | 1.8 | 1.6 | 1.1 |
| 625478 | $MnO_2$/seeding | 3.3 | 4.7 | 2.7 |
| 625540 | $MnSO_4$/seeding | 1.3 | 1.1 | 0.5 |
| 625601 | $MnSO_4$/no seeding | 2.5 | 1.8 | 1.6 |
| 625653 | $MnO_2$/no seeding | 2.5 | 1.6 | 1.3 |
| 625668 | $MnO_2$/no seeding | 0 | 0 | 0 |
| 625694 | $Mn_2O_3$/no seeding | 3.1 | 0.7 | 0.7 |
| 626000 | MnO/seeding | 7.4 | 5.8 | 5.3 |

**Table S2.**

Probability of occurrence of strong shots, fringes, and fringes that are also strong shots. Each run has 551 shots. Strong shots are defined as having photon counts with more the 107 detector units. Fringes are identified using our empirical rules presented in the previous section.

| State No. | Group | Emission Energy (eV) | Transition Dipole (a. u.) | Inverse Radiative Lifetime (a. u.) |
|---|---|---|---|---|
| 1 | K$\alpha_1$ | 5902.3 | 0.008707 | 0.004040 |
| 2 | K$\alpha_1$ | 5901.1 | 0.007426 | 0.002938 |
| 3 | K$\alpha_1$ | 5900.1 | 0.005651 | 0.001702 |
| 4 | K$\alpha_1$ | 5890.3 | 0.004566 | 0.001111 |
| 5 | K$\alpha_1$ | 5899.6 | 0.003958 | 0.0008347 |
| 6 | K$\alpha_2$ | 5889.1 | 0.003836 | 0.0007840 |
| 7 | K$\alpha_2$ | 5889.1 | 0.003590 | 0.0006869 |
| 8 | K$\alpha_2$ | 5890.2 | 0.002587 | 0.0003565 |
| 9 | K$\alpha_2$ | 5889.1 | 0.002340 | 0.0002917 |
| 10 | K$\alpha_2$ | 5889.0 | 0.002035 | 0.0002206 |
| 11 | K$\alpha_2$ | 5890.5 | 0.001728 | 0.0001591 |
| 12 | K$\alpha_1$ | 5898.7 | 0.001406 | 0.0001054 |

**Table S3.**

Energies, transition dipoles and inverse radiative lifetime ($\Gamma$rad) of final states considered in the simulation.